# Scattering due to non-magnetic disorder in 2D anisotropic d-wave high $T_c$ superconductors

**P. Contreras**[*], **Dianela Osorio**

Department of Physics, University of Los Andes, Mérida 5101, Venezuela

[*]Corresponding author (pcontreras@ula.ve)

**Abstract:** Inspired by the studies on the influence of transition metal impurities in high Tc superconductors and what is already known about nonmagnetic suppression of Tc in unconventional superconductors, we set out to investigate the behavior of the nonmagnetic disordered elastic scattering for a realistic 2D anisotropic high Tc superconductor with line nodes and a Fermi surface in the tight-binding approximation. For this purpose, we performed a detailed self-consistent 2D numerical study of the disordered averaged scattering matrix with nonmagnetic impurities and a singlet line nodes order parameter, varying the concentration and the strength of the impurities potential in the Born, intermediate and unitary limits. In a high Tc anisotropic superconductor with a tight binding dispersion law averaging over the Fermi surface, including hopping parameters and an order parameter in agreement with experimental data, the tight-binding approximation reflects the anisotropic effects. In this study, we also included a detailed visualization of the behavior of the scattering matrix with different sets of physical parameters involved in the nonmagnetic disorder, which allowed us to model the dressed scattering behavior in different regimes for very low and high energies. With this study, we demonstrate that the scattering elastic matrix is affected by the non-magnetic disorder, as well as the importance of an order parameter and a Fermi surface in agreement with experiments when studying this effect in unconventional superconductors.
Keywords: Unconventional Anisotropic Superconductors, Lifetime, Non Magnetic Disorder, Unitary, Intermediate and Born Regimes

**Keywords:** Unconventional Anisotropic Superconductors, Lifetime, Non Magnetic Disorder, Unitary, Intermediate and Born Regimes.

## 1. Introduction

The discovery of high temperature superconductivity led to a new era in solid state physics. It was a ceramic composite known as calcium doped lanthanum cuprate having a transition temperature $T_c$ of 30 K (Bednorz and Muller 1986).

This $T_c$ was already high to suggest that it might be difficult to explain the phenomenon using the BCS Theory. The next year it was found a close related material with a $T_c$ of about 93 K (Wu et al. 1987). High $T_c$ cuprate superconductors (HTSCs) are anisotropic layered structures, containing $CuO_2$ planes. All of them have lots of amazing properties, that cannot be explained by the BCS theory of superconductivity (Sheadem 1994, Waldran 1996, Cava 2000).

In particular, to this work, the first experiments investigating the influence of transition metal impurities in HTSCs, showed that nonmagnetic disorder suppress superconductivity, more strongly than magnetic impurities disorder (Gang et al. 1987, Momono et al 1994, Sanikidze et al. 2005), on the contrary to BCS superconductors, where only magnetic impurities reduce the transition temperature ($T_c$).

It was suggested (Sun and Maki 1995) that Momono and collaborators experiments showed that the $La_{2-x}Sr_xCuO_4$ compound was in the unitary scattering regime. Sometime later it was observed that in disordered $La_{2-x}Sr_xCuO_4$ doped with non-magnetic Sr impurities (Sr is an alkali earth metal), the transition temperature $T_c$ started to decrease rapidly (Momono et al. 1996). Nowadays, low temperatures properties of superconducting doped $La_{2-x}Sr_xCuO_4$ continue to be intensively investigated (Dalakova et al. 2018).

On the other hand, it has been proposed that in HTSCs, the superconducting gap corresponds to a paired singlet state $\Delta(k) = \Delta(-k)$ and with certain kind of nodes. In particular, one of these gaps has $d_{x^2-y^2}$ symmetry (Scalapino 1995). The superconducting gap for this symmetry has lines nodes on the Fermi surface corresponding to the one-dimensional irreducible representation $B_{1g}$ of the tetragonal point symmetry group $D_{4h}$ (Tsuei and Kirtley 2000). In a $d_{x^2-y^2}$ symmetry gap, nonmagnetic disorder strongly quenches superconducting ordering leading to strong suppression of $T_c$ (Sanikidze et al. 2005).

In this work, we use a tight binding anisotropic model with a nearest neighbor expression for a band sheet centred at the corners (±π/a, ±π/a) of the first Brillouin zone in 2D, $\xi(k_x, k_y) = \epsilon +$

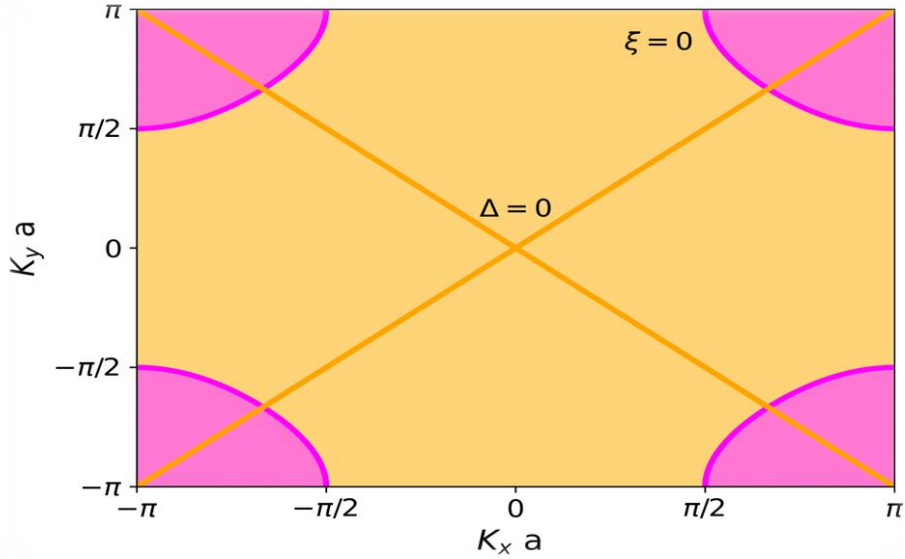

$2t[\cos(k_x a) + \cos(k_y a)]$ with well-established from experiments hopping parameters $(t, \epsilon)$ = (0.2, 0.4) meV and an electron-hole (eh) symmetry $\xi_e(k_x, k_y) = \xi_h(-k_x, -k_y)$ (Yoshida et al. 2012).

For the k dependence of the gap, we use a realistic 2D tight binding expression corresponding to lines nodes in the Fermi surface $\Delta(k_x, k_y) = \Delta_0 \phi(k_x, k_y)$, with $\phi(k_x, k_y) = [\cos(k_x a) - \cos(k_y a)]$ and

*Figure 1. 2D implicit plot of the tight binding anisotropic Fermi surface and superconducting gap with line nodes.*

$\Delta_0$ = 33.9 meV, which is in agreement with optimally doped (x = 0.15) $La_{2-x}Sr_xCuO_4$ experiments (Yoshida et al. 2012). We follow (Tsuei et al. 1997) arguments and use a pure $d_{x^2-y^2}$ order-parameter symmetry to model HTSCs. This expression has been used recently to study the doping dependence of the pairing symmetry in $La_{2-x}Sr_xCuO_4$ (Verma et al. 2019, Gupta et al. 2019).

Figure 1 shows the implicit plots for the Fermi surface $\xi_F(k_x, k_y) = 0$ (red violet shadowed region) and the line gap nodes $\Delta(k_x, k_y) = 0$ (orange lines) for the set of tight binding parameters of the previous paragraph. The four intersections of the Gap and the Fermi surface contain the nodal quasiparticles region that is modeled at low frequencies.

Our contribution to this work is the study the impurity scattering disorder in a self-consistent manner for $\widetilde{\omega}(\omega + i\,0^+)$ with a tight binding anisotropic model in 2D. TB anisotropic modeling allowed us to successfully fit experimental low temperature data in another unconventional multiband superconductor at very low temperatures (Contreras et al. 2004, Contreras 2011).

The computational and mathematical details of the algorithm were tested and reported for an isotropic Fermi surface and its corresponding order parameter in a previous work (Contreras and Moreno, 2019).

The structure of this paper is as follows. In section 2, we introduce the theoretical formalism of the elastic scattering non-magnetic disordered averaged matrix $\widetilde{\omega}(\omega + i\, 0^+)$. In section 3 we model the behavior of the imaginary part of the electron-hole symmetric scattering matrix depending on the disorder parameters, i.e., the impurity disorder concentration $\Gamma^+$, and the inverse of the strength of the impurities potential c in the Born, intermediate, and unitary regimes.

The inverse of the imaginary part of the scattering matrix $\Im^{-1}[\widetilde{\omega}(\omega + i\, 0^+)]$ enters the expressions for the kinetic coefficients of unconventional superconductors at very low energies and temperatures (Pethick and Pines 1986), that is, the thermal conductivity, the sound attenuation, and the electrical resistivity.

The universal behavior has been observed experimentally and studied theoretically at very low temperatures (see Mineev and Samokhin 1999 and references therein for a review of non-magnetic disorder in unconventional superconductors).

For reviews on the question, whether a gap with nodes can explain experimental data on anisotropic HTSCs, please see (Walker 2000, Bozovic et al. 2018, Shaginyan 2019). For a discussion of other impurity effects affecting HTSCs and other unconventional superconductors, see (Pogorelov and Loktev, 2018). Finally, for Fermi and Bose atomic gases at ultra-cold temperatures there is a study emphasizing the importance of the unitary limit and the scattering matrix analysis (Pitaevskii 2008).

## 2. Formalism for non-magnetic impurities disorder

In this section, we present the main equations for the scattering formalism involving the self-energy matrix $\widetilde{\omega}(\omega + i\, 0^+)$ in the case of non-magnetic disorder (Schachinger and Carbotte 2003) following a realistic approach for modelling experimental data.

This approach was introduced to calculate the residual absorption at zero temperature in d-wave superconductors (Schachinger and Carbotte 2003, Carbotte and Schachinger 2004). The formalism comes from the combination of the Gorkov Greens function (Gorkov 1958, Abrikosov et al. 1963) and the Edwards non-magnetic disorder scattering techniques (Edwards 1962, Ziman 1979).

This formalism has been used to study the low temperature behavior of several physical kinetic properties in Heavy Fermions, Ruthenates, and HTSCs among other unconventional superconductors (Pethick and Pines 1986, Schachinger and Carbotte 2003, Schuerrer et al. 1999, Carbotte and Schachinger 2004, Hirschfeld et al. 1988, Balatsky et al. 2006, Mineev and Samokhin 1999), it allows fitting experimental low temperature data in the unitary region, where the Boltzmann equation approach does not work.

The equation for $\widetilde{\omega}(\omega + i\, 0^+)$ can be written in the following way (Schachinger and Carbotte 2003)

$$\widetilde{\omega}(\omega + i\, 0^+) = \omega + i\, \pi\, \Gamma^+ \frac{g(\widetilde{\omega})}{c^2 + g^2(\widetilde{\omega})} \quad (1)$$

The function $\widetilde{\omega}(\omega + i\, 0^+)$ in this particular case, describes the self-consistent renormalization of the quasiparticle energy (dressed frequencies $\widetilde{\omega}$) due to elastic impurity scattering on non-magnetic impurities disorder in the case of electron-hole symmetry, i.e. $g(\widetilde{\omega}) = g_0(\widetilde{\omega})$.

Following (Schachinger and Carbotte 2003), the parameter $c = {1}/{(\pi\, N_F\, U_0)}$ represents the inverse of the impurities strength with $N_F$ is the density of states at the Fermi surface, and $U_0$ is the impurity potential. The parameter $\Gamma^+ = {n_{imp}}/{(\pi^2\, N_F)}$ is proportional to the impurity concentration $n_{imp}$.

Non-magnetic scattering disorder in normal metals assumes the following physical conditions (Edwards 1962, Ziman 1979): there are N impurities equal and independent of each other which create random disorder (on a macroscopic scale the crystal is homogeneous), and the impurities scatter electrons elastically (i.e. there is no energy loss in collisions) following quantum mechanics scattering rules (Landau and Lifshitz, 1981).

The function $g(\widetilde{\omega})$ in (1) is given by the following expression

$$g(\widetilde{\omega}) = \langle \frac{\widetilde{\omega}}{\sqrt{\widetilde{\omega}^2 - \Delta^2(k_x, k_y)}} \rangle_{FS} \quad (2)$$

The average $\langle ... \rangle_{FS}$ is performed over the tight binding 2D Fermi surface "$\xi(k_x, k_y)$" according a technique successfully used to fit experimental data on the low T superconducting sound attenuation and electronic thermal conductivity (Contreras et al. 2004, Contreras 2011).

Finally, if electron-hole symmetry (eh) is not considered, the other spin Pauli components of $g(\widetilde{\omega})$ have to be taken into consideration, that is, $g_1(\widetilde{\omega})$ and $g_3(\widetilde{\omega})$, but for a d-wave HTSCs with a singlet gap $g_1(\widetilde{\omega}) \sim \langle \Delta(k_x, k_y) \rangle_{FS} = 0$, and $g_3(\widetilde{\omega}) \sim \langle \xi(k_x, k_y) \rangle_{FS} = 0$ follows from eh symmetry (Bang 2012).

Born's approximation applies and if c >> 1 (i.e. $U_0$ << 1) with a new disorder parameter $\Gamma_B^+ = \Gamma^+/c^2$ << 1. However, a realistic value for the inverse of the strength is $c \sim 0.4$ for reasonable numbers for the disorder concentration $\Gamma^+$ in the Born limit as we will see from the simulation. In the Born limit (1) becomes

$$\widetilde{\omega}(\omega + i\, 0^+) = \omega + i\, \pi\, \Gamma_B^+\, g(\widetilde{\omega}) \quad (3)$$

In addition for large values of $U_0$, c $\to 0$ and this limit represents the unitary regime which is given by equation

$$\widetilde{\omega}(\omega + i\, 0^+) = \omega + i\, \pi\, \Gamma^+ \frac{1}{g(\widetilde{\omega})} \quad (4)$$

The imaginary part of (1) defines the inverse of the quasiparticle disordered averaged lifetime as

$$\frac{1}{\tau(\omega)} = 2\, \Im[\widetilde{\omega}(\omega)] \quad (5)$$

For very low frequencies when $\omega \to 0$, $\widetilde{\omega} = i\gamma$. The quantity $\gamma$ defines the "zero dressed" or "impurity averaged" frequency of the zero energy effective elastic scattering rate (Schachinger and Carbotte 2003) in the superconducting state as the function $\gamma(\Gamma^+, c)$ with $\gamma(\Gamma^+, c) = \pi\Gamma^+ g(i\gamma)/[c^2 + g^2(i\gamma)]$ and $g(i\gamma) = \langle \gamma/\sqrt{\gamma^2 + |\Delta|^2} \rangle_F$.

$\gamma(\Gamma^+, c)$ is the transcendental equation of the residual impurity averaged lifetime at zero frequency ($\omega = 0$) with a residual disordered averaged lifetime define as $\gamma = 1/\tau(0)$, it has been study in (Schachinger and Carbotte 2003, Carbotte and Schachinger 2004, Contreras and Moreno, 2019)

$\gamma(\Gamma^+, c)$ determines the crossover energy scale separating the two scattering limits as we will see in the following section. If the energy of excitations is greater than $\gamma$ then self-consistence can be neglected and we use the Boltzmann equation for calculating the kinetic properties in HTSCs (Arfi and Pethick 1988), but if the typical energy is smaller than $\gamma$, self-consistency cannot be neglected in the physical kinetics at low energies. For a 2D line nodes HTSC order parameter in the Born region it is found that

$\gamma_B \sim e^{-\pi c_0}$ with $c_0 = \tau_n \Delta_0$ and in the unitary region $\gamma_U \sim \Delta_0 \sqrt{0.5\,\pi/(c_0 \ln c_0)}$ (Walker et al, 2001).

To conclude this section, the energy uncertainty principle allows the low energy quasiparticles to have a spread in energy of the order of $\Gamma^+$, henceforth disordered dressed quasiparticles $\widetilde{\omega} \sim \omega + i\,\Gamma^+$ have a lifetime $\tau \sim 0.5/\Gamma^+$. In addition, the superconducting density of states (DOS) for line nodes HTSCs in the Born limit is approximated $N_B/N_F \sim \omega/\Delta_0$ and in the unitary limit is approximated by $N_U/N_F \sim \Gamma^+/\Delta_0$, this derivations shows the existence of normal states dressed quasiparticles at zero energy in the unitary limit.

## 3. Numerical results and $\widetilde{\omega}\,(\omega + i\,0^+)$ visualization

### 3.1. Evolution from the unitary regime to the Born limit.

In this subsection, we model the solution of (1) by varying the parameter strength "c", that is, $c = 1/(\pi\,N_F\,U_0)$, and by fixing the value of disorder concentration $\Gamma^+$ for two cases of physical interest. The first case is for a value of $\Gamma^+ = 0.15\ meV$, which resembles optimally doped values of impurities in experimental samples. The second case is for a very dilute disorder concentration $n_{imp}$, that is, $\Gamma^+ = 0.01\ meV$.

Figure 2 shows the evolution of $\Im[\widetilde{\omega}]\,(\omega + i\,0^+)$ as function of the parameter c which is inversely proportional to the strength of the impurities $U_0$ from values $U_o \gg 1\ and\ c \xrightarrow{+} 0$ (violet curve) to values $U_o \ll 1\ and\ c \xrightarrow{-} 1$ (yellow curve) i. e. when $\pi\,N_F\,U_0 \to 1$, $(N_F \sim 10^{15}eV^{-1})$ for an electron-hole symmetric tight binding dispersion and for a disorder concentration of $\Gamma^+ = 0.15$ meV.

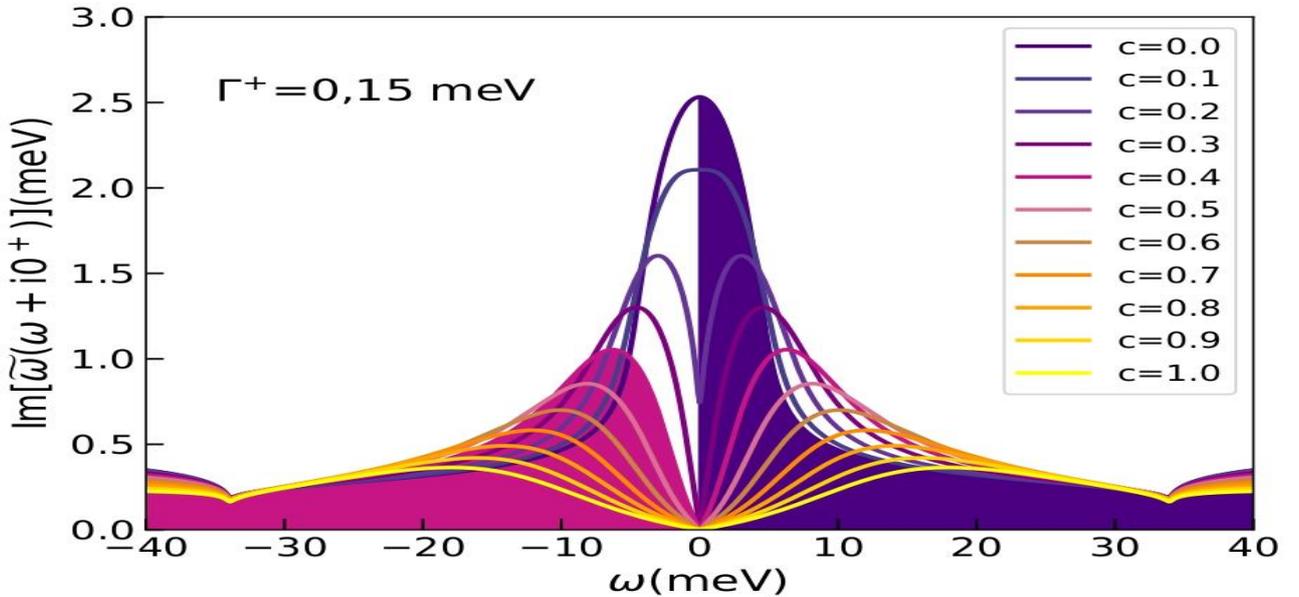

**Figure 2.** *Plot of the Im $\widetilde{\omega}(\omega + i\,0^+)$ as a function of the inverse scattering strength c for an optimal disorder concentration.*

The region under $\Im[\widetilde{\omega}]\,(\omega + i\,0^+)$ in the unitary limit corresponding to c = 0.0 (violet curve) has been shaded violet in the right side of the plot. On the other hand, a well define Born scattering region with a smooth minimum at zero frequencies corresponds to an inverse strength of c = 0.4 according our calculation. In the left side of fig 2, the Born region under $\Im[\widetilde{\omega}]\,(\omega + i\,0^+)$ has been shaded red. In figure 2, disorder affects most strongly the low energy region near up to, say, 5 meV.

For the unitary limit (violet shaded region) in the right side of fig 2, we observe a maximum centred at zero frequencies ($\omega = 0$) in the $\Im[\widetilde{\omega}]\,(\omega + i\,0^+)$ function which is clearly an indication of strong disorder. It corresponds to a minimum in the scattering lifetime of the dressed quasiparticles (which are unitarily bounded to the non-magnetic disorder potential $U_0$).

In figure 2, for $\Im_U[\widetilde{\omega}]$ we do not observe a minimum, instead we see a flattening of the function for higher energies (monotonically drops as frequency increases), it indicates a constant lifetime in the unitary limit for the normal state.

For the Born shaded red region in the left side (negative frequencies) of fig 2, we observe a maximum $max\,\Im_B[\widetilde{\omega}]$ scattering centred at higher non-zero frequencies in the $\Im[\widetilde{\omega}]\,(\omega + i\,0^+)$ function (a maximum in scattering rate occurs around ω = 4 meV for an optimal disorder value of $\Gamma^+$=0.15 meV, which is a larger value than its value at ω = 0 meV denoted by γ).

That maximum in scattering corresponds to a minimum in the lifetime of Born (weakly bounded to the non-magnetic impurities disorder potential $U_0$) quasiparticles. We also observe a displacement of the $max\,\Im_B[\widetilde{\omega}]$ as the inverse of the strength c increases. A weak scattering potential blurs $max\,\Im_B[\widetilde{\omega}]$, spreading the width of the peak. We also observe a smooth minimum at frequencies γ = 0 meV for most clear colors lines (from c = 0.4 to c = 1 values).

At energies $\omega = \Delta_0 \sim \pm 33.9\,meV$, we observe the transition in $\Im[\widetilde{\omega}]\,(\omega + i\,0^+)$ from the superconducting to normal state as an small abrupt change in the slope of the function $\Im[\widetilde{\omega}]$ for both regimes. In the normal state the function $\Im[\widetilde{\omega}]$ tends to be constant and slight varies for different initial residual $\gamma$ values, indicating a constant quasiparticles disordered lifetime in the normal state.

Therefore, we define the unitary nonmagnetic disordered averaged quasiparticles, as the ones that have a smooth $max\,\Im_U[\widetilde{\omega}]$ given by the value $\gamma\,(\Gamma^+, c) = \gamma\,(0.15, 0)$ for the case of eh symmetry.

We also define the Born nonmagnetic disordered average quasiparticles, as the ones that have an smooth $max\,\Im_B[\widetilde{\omega}]$ given by the value$s\,(\Gamma^+, c) = (0.15, 0.4)$ and that have a smooth minimum given by the value $\gamma\,(\Gamma^+, c) = 0$ for the case of eh symmetry.

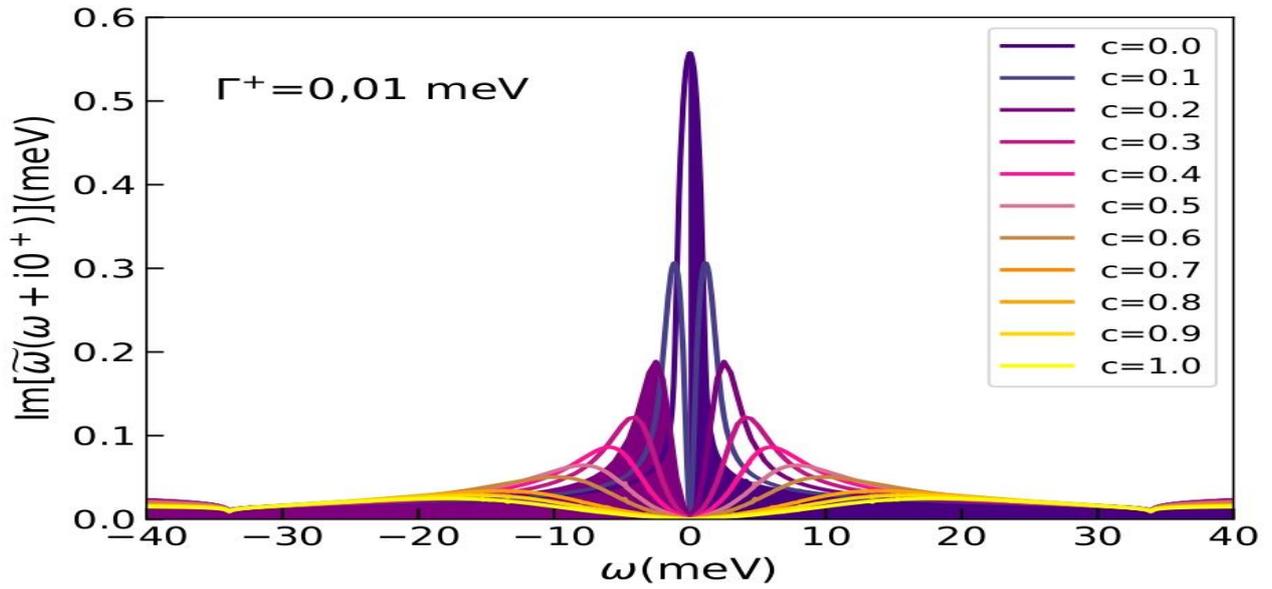

***Figure 3.*** *Plot of $Im\ \tilde{\omega}(\omega +i\ 0^+)$ as a function of the inverse scattering strength c for a very dilute disorder.*

Figure 3 shows the evolution of $\Im[\tilde{\omega}](\omega + i\ 0^+)$ as a function of c for a dilute concentration of impurities, that is, $\Gamma^+ = 0.01$ meV.

The region for $\Im[\tilde{\omega}](\omega + i\ 0^+)$ in the unitary limit corresponding to c = 0.0 has been shaded violet in the right side of the plot as in the previous case, we observe a much smaller area compared with the previous optimal disorder case. On the other hand, a well define Born scattering region corresponds now to a smaller c = 0.2 value. In the left side of fig 3 for negative frequencies, the Born region has been shaded saturated violet, that indicates that for very diluted $\Gamma^+ = 0.01\ meV$ most of the behavior correspond to the Born case, as we expected for the analysis in section 2.

For the unitary violet region in the right side of fig 3, we observe a maximum centred at zero frequencies in the $\Im[\tilde{\omega}](\omega + i\ 0^+)$ function with a much smaller value of zero residual frequencies $\gamma$. For $\Im_U[\tilde{\omega}]$ we do not observe a minimum, but instead as in the previous case, we see a monotonically flattening of the function for higher energies, that indicates a constant normal state quasiparticles lifetime in the unitary limit for the case of very dilute impurities concentration.

For the Born saturated violet region in the left side of fig 3, we observe a $max\ \Im_B[\tilde{\omega}]$ centred at non zero frequencies in the $\Im[\tilde{\omega}](\omega + i\ 0^+)$ function starting from c = 0.2, it corresponds to a minimum in the quasiparticle scattering lifetime for dilute disorder. We also observe the displacement of the $max\ \Im_B[\tilde{\omega}]$ as c increases. As before, a weak scattering potential blurs the maximum, spreading monotonically the width of the peak and decreasing the $max\ \Im_B[\tilde{\omega}]$. In figure 3, for the function $\Im[\tilde{\omega}](\omega + i\ 0^+)$ we observe a pronounced minimum at zero frequencies with a exponentially increasing frequency dependence starting with the c = 0.1 value (blue color).

At energies $\omega = \Delta_0 \sim \pm 33.9\ meV$, we also observe the transition in $\Im[\tilde{\omega}](\omega + i\ 0^+)$ from the superconducting to normal state as an small abrupt change in the slope of the function $\Im[\tilde{\omega}]$ for both regimes and all c values. In table 1, we summarize our findings for figs 2 and 3.

Table 1: Disorder parameters dependency for very dilute and optimal concentration of non-magnetic impurities.

| Disorder parameters: the strength c and the concentration $\Gamma^+$. | Very dilute impurities concentration: $\Gamma^+$ = 0.01 meV | Optimal impurities concentration: $\Gamma^+$ = 0.15 meV |
|---|---|---|
| c = 0 ($U_0 \gg 1$) Strong disorder (violet line). | Unitary behavior, with a maximum $\Im_U[\widetilde{\omega}]$ and $\gamma < 1$. | Unitary behavior, with a maximum $\Im_U[\widetilde{\omega}]$ and $\gamma > 1$. |
| c = 0.2 Intermedia disorder (saturated violet) | Born behavior with two displaced maximums $\Im_B[\widetilde{\omega}]$ and one minimum at zero frequencies. | Undefined behavior with a minimum at zero frequencies but maximum at higher frequencies. |
| c = 0.4 Weak disorder (red line) | Born behavior with two displaced maximums $\Im_B[\widetilde{\omega}]$ and one minimum at zero frequencies. | Born behavior with two displaced maximums $\Im_B[\widetilde{\omega}]$ and one minimum at zero frequencies |
| c = 1 ($U_0 \ll 1$) Very weak disorder (yellow line). | Born extended (almost constant behavior). | Born extended (almost constant behavior). |

## 3.2. Disorder evolution inside the unitary, the Born and the intermedia limits.

In this subsection, we firstly study the behavior of $\Im[\widetilde{\omega}] (\omega + i\, 0^+)$ for the unitary strength limit c = 0 and varying impurities concentration, from values of $\Gamma^+$ starting at very diluted disorder (yellow line), dilute disorder (orange line), an almost optimal disorder (brown line), an optimal disorder (red line), and finally an enriched disorder (violet line).

Figure 4 shows the evolution of $\Im[\widetilde{\omega}] (\omega + i\, 0^+)$ at c = 0 for the five values of $\Gamma^+$ in meV. We observe the smooth maximum centred at zero frequencies $\gamma$ in $\Im_U[\widetilde{\omega}]$ with a much smaller value of residual $\gamma$ for very dilute values of disorder $\Gamma^+$.

For $\Im_U[\widetilde{\omega}]$ we do not observe a minimum, but instead as in previous cases, we see a monotonically flattening of the function for higher energies indicating a constant lifetime value in the unitary limit for the normal state for all cases of non-magnetic disorder. We observe that the value of $\Im_U[\widetilde{\omega}]$ in the normal state (above 33.9 meV) depends on the disorder concentration, given the optimal doped disorder a value of 0.40 meV, meanwhile the very dilute disorder gives 0.01 meV.

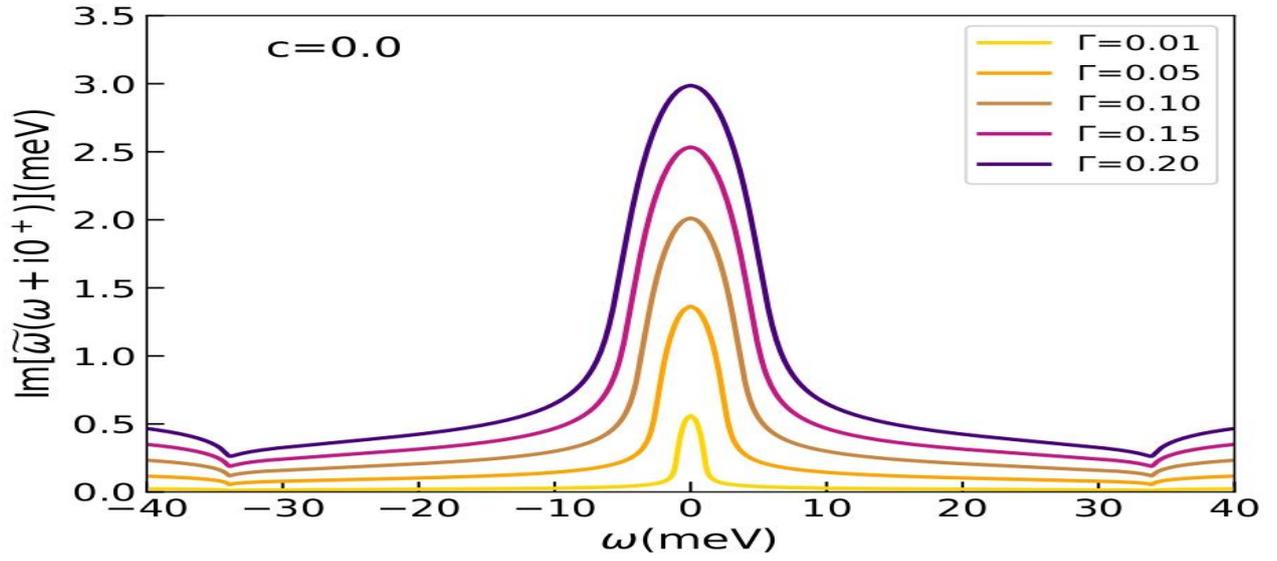

**Figure 4.** *Plot of the tight binding Im $\widetilde{\omega}(\omega + i0^+)$ as a function of the disorder averaged impurities concentration $\Gamma^+(meV)$ for an strong impurities potential $U_0$ in the unitary limit.*

Figure 5 shows the evolution of $\Im[\widetilde{\omega}](\omega + i\,0^+)$ at c = 0.4 (Born scattering limit) as function of the five values of disorder $\Gamma^+$.

We observe a $max\ \Im_B[\widetilde{\omega}]$ centred at $\omega \sim \pm 4\ meV$ in the function for electrons and holes. A very dilute disorder $\Gamma^+$ blurs the maximum, spreading monotonically the width of the peak and decreasing the $max\ \Im_B[\widetilde{\omega}]$. For the function $\Im_B[\widetilde{\omega}]$ we observe a minimum at zero disordered averaged frequencies. For very diluted disorder $\Gamma^+$ we observe a power increase at very low energies, and for optimal disorder we see an exponential increase of $\Im_B[\widetilde{\omega}]$ at very low energies.

We notice that the value of $\Im_B[\widetilde{\omega}]$ in the normal state (above 33.9 meV) also depends on the disorder concentration, given the optimal doped disorder a value of 0.50 meV, meanwhile the very dilute disorder gives 0.01 meV.

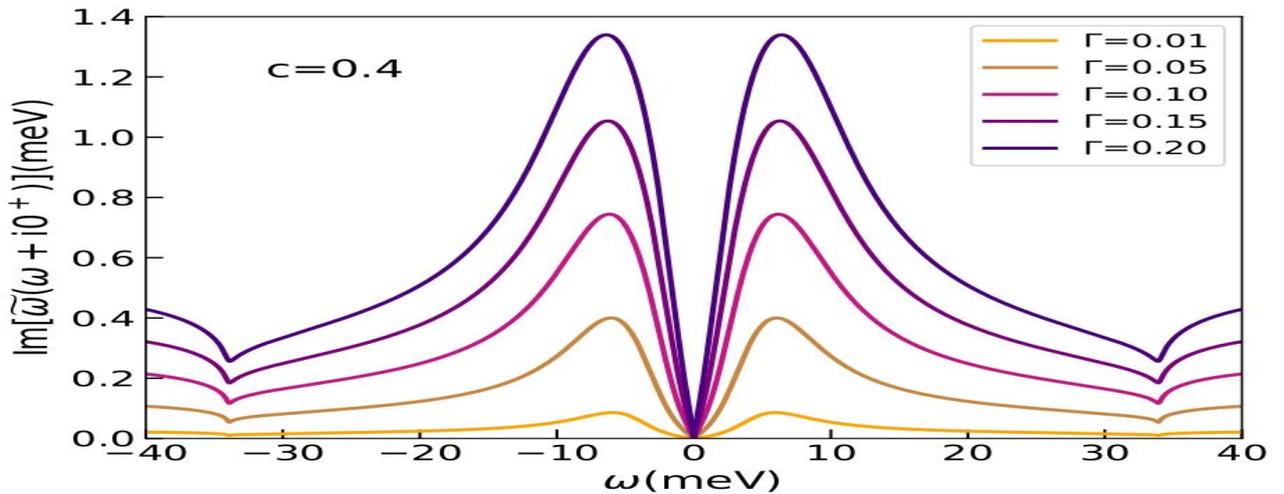

**Figure 5.** *2D Plot of the tight binding Im $\widetilde{\omega}(\omega + i0^+)$ as a function of the disorder averaged concentration $\Gamma^+(meV)$ for the weak disorder potential $U_0$ in the Born limit region.*

To complete the whole picture, figure 6 shows the disorder $\Gamma^+$ evolution of $\Im_I[\tilde{\omega}](\omega + i\, 0^+)$ at c = 0.2 which is an intermedia region between Born and unitary limits.

We observe a different from zero min $\Im_I[\tilde{\omega}](0)$ centred at zero frequencies $\gamma$ in the $\Im_I[\tilde{\omega}]$ function as happens in the Born scattering case. Small values of the zero residual frequency $\gamma$ appear for very dilute values of $\Gamma^+$ in the intermedium case

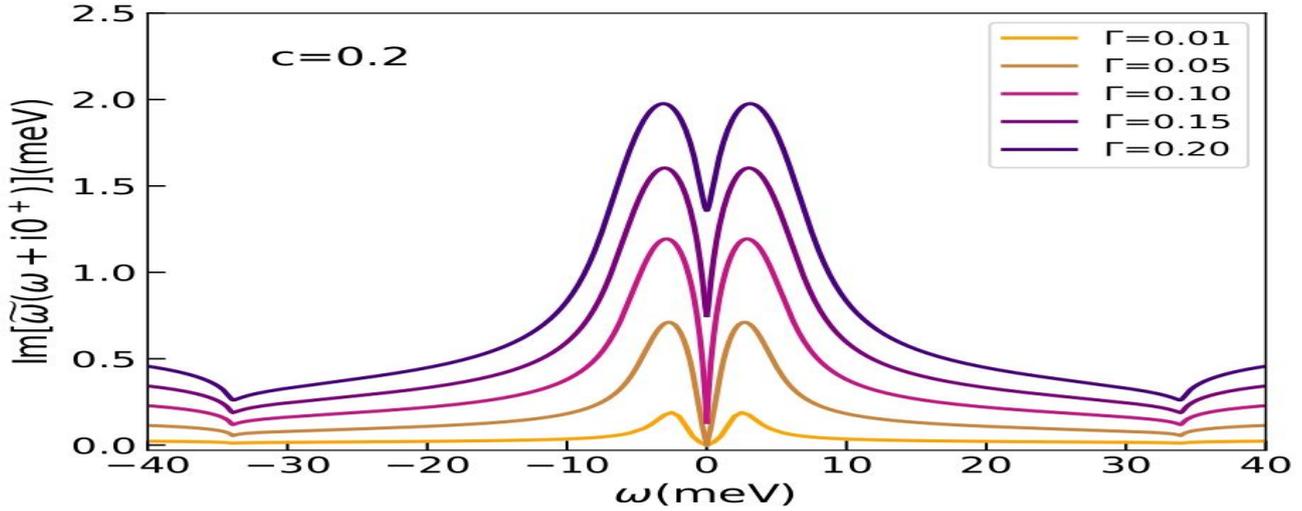

**Figure 6.** *Plot of the tight binding Im $\tilde{\omega}(\omega + i\, 0^+)$ as a function of the disorder averaged impurities concentration $\Gamma^+(meV)$ for intermedia values of the disorder potential $U_0$ (c = 0.2).*

We also observe a $max\, \Im_I[\tilde{\omega}]$ centred at $\omega \sim \pm 4\, meV$ in the $\Im_I[\tilde{\omega}](\omega + i\, 0^+)$ function as for the Born limit. In this case there will be finite $\gamma$ values for all disorder concentrations as happens in the unitary case. From fig. 6, we observe an exponential increase of $\Im_I[\tilde{\omega}]$ at very low energies for all $\Gamma^+$.

Finally, from figures 4, 5, and 6, we observe that the $max\, \Im_{U,B,I}[\tilde{\omega}]$ for the three functions is different in values for all $\Gamma^+$. For the unitary case, the maximum value (3.0 meV) doubles the value for maximum in the Born case (1.3 meV), meanwhile in the intermedium case has a max. value of 2.0 meV.

## 4. Conclusions

The present work was aimed at investigating the behavior of the elastic scattering non-magnetic disordered averaged matrix $\tilde{\omega}(\omega + i\, 0^+)$ for a realistic 2D anisotropic HTSC with line nodes and a tight binding Fermi surface and gap.

The results are summarized in two subsections of section 3. In subsection 3.1, we modeled first the behavior of the imaginary part of the electron-hole symmetric scattering matrix depending on eleven values of c (the inverse of the strength of the disorder potential $U_0$) for two disorder regions of physical importance. First, an optimal disorder region with $\Gamma^+$ = 0.15 meV and second for a very dilute region with $\Gamma^+$ = 0.01 meV. The results were visualized in figures 2 and 3 and a summary of the results is given in table 1.

Subsection 3.1 visualizes the behavior of the disordered matrix $\tilde{\omega}(\omega + i\, 0^+)$ inside the unitary (c = 0) intermedia (c = 0.2) and Born (c= 0.4) regions for five disorder concentrations, starting at very diluted disorder, dilute disorder, an almost optimal disorder, an optimal disorder, and finally an enriched disorder. The results were visualized in figures 4, 5 and 6 and the analysis with the results is given in subsection 3.2. We found in this section that the evolution of the disordered matrix $\widetilde{\omega}(\omega + i\,0^+)$ depends strongly on the value of $\Gamma^+$.

## Acknowledgements

P. Contreras thanks Dr. Yu. Beliayev for stimulating discussions that helped to clarify some experimental aspects of non-magnetic disorder scattering in HTSCs. The authors did not receive financial support for research, authorship and/or publication of this article.